\documentclass{aa}
\usepackage{times}
\usepackage{graphicx}
\usepackage{natbib}

\bibpunct{(}{)}{;}{a}{}{,}

\begin{document}

\def\asca{{\it ASCA}~}
\def\be{\begin{equation}}
\def\bhm{M_{\rm BH}}
\def\chandra{{\it Chandra~}}
\def\cj{{\cal J}}
\def\dotm{\dot{m}}
\def\ee{\end{equation}}
\def\fnu{f_{\nu}}
\def\fth{f_{\rm th}}
\def\ii{\vec{i}}
\def\jj{\vec{j}}
\def\kk{\vec{k}}
\def\nd{\vec{n}_{\rm d}}
\def\no{\vec{n}_{\rm o}}
\def\rin{r_{\rm in}}
\def\rout{r_{\rm out}}
\def\sunm{M_{\odot}}
\def\II{I_{\nu}(\mu,\tau_{\nu})}
\def\SS{S_{\nu}(\mu,\tau_{\nu})}
\def\DM{\dot{M}}
\def\OME{\Omega}
\def\QA{Q_{\rm adv}}
\def\QV{Q_{\rm vis}}
\def\QR{Q_{\rm rad}}
\def\Qdadv{Q_{\rm adv}^{\rm d}}
\def\Qdcool{Q_{\rm cool}^{\rm d}}
\def\Qc{Q_{\rm c}}
\def\Qcadv{Q_{\rm adv}^{\rm c}}
\def\Qd{Q_{\rm vis}}
\def\sax{{\it BeppoSAX~}}
\def\tr{t_{r\phi}}
\def\ts{T_{\rm s}}
\def\dotm{\dot{m}}
\def\dotM{\dot{M}}
\def\dm{\dot{M}}
\def\Ome{\Omega}
\def\XMM{{\it XMM}}

\def\hd{H_{\rm d}}
\def\vrd{v_{\rm r}^{\rm d}}
\def\rhod{\rho_{\rm d}}
\def\pd{P_{\rm d}}

\title{Extreme Slim Accretion Disks and Narrow Line Seyfert 1 Galaxies:
the Nature of the soft X-ray Hump}

\author{Jian-Min Wang\inst{1,2,3}
   \and Hagai Netzer\inst{1}}

\titlerunning{Nature of the Soft Humps in NLS1}
\authorrunning{J.-M. Wang and H. Netzer}

\offprints{J.-M. Wang,\\ \email{wang@astro.uni-tuebingen.de}}

\institute{School of Physics and Astronomy and the Wise Observatory,
Tel Aviv University, Tel Aviv 69978, Israel
\and Laboratory of High Energy Astrophysics, Institute of High Energy 
Physics, Chinese Academy of Sciences, Beijing 100039, P.R., China
\and Institut f\"ur Astronomie und Astrophysik, Abt. Astronomie,
Universit\"at T\"ubingen, Sand 1, D-72076 T\"ubingen, Germany} 

\date{Received $<$date$>$ / Accepted $<$date$>$ }

\abstract{We present a detailed  model of an extreme slim disk (ESD) with a 
hot corona around a massive black hole with 
dimensionless accretion rate $\dotm$  in the range $2.5\ll \dotm\leq 100$.
We assume that a fraction $f$ of the gravitational energy is released in the
hot corona and the rest is released in the ESD. The energy equation of the
ESD is dominated by advection and the spectrum shows a broad ``hump'' caused 
by saturated Comptonization with monochromatic luminosity  given by  
$\nu L_{\nu}\propto \nu^0$. This relationship enables us to estimate the black
 hole mass from the ESD
luminosity. 
The spectrum of the
hot corona is sensitive to the parameter $f\dotM$ and the viscosity and shows
a Comptonized  power-law with a high-energy  cutoff. 
The model is used  to
explain the spectral properties of narrow line  Seyfert 1 galaxies (NLS1s). 
In particular, it can
 explain the spectrum of extreme NLS1s like NLS1 RE J1034+396. Our spectral
estimate of the black hole mass in this source is in good agreement with the
mass obtained by applying the reververation mapping correlation.
 We also
examine the Eddington ratios in a large NLS1 sample and find that most objects
show super-Eddington accretion rates.  We argue that soft X-ray humps in
NLS1s are natural consequences of super-Eddington accretion in such objects.
}

\maketitle

\section{Introduction}\label{sect:intro}
Narrow line Seyfert 1 galaxies (NLS1) receive much attention because of
their unusual  properties: the H$\beta$ line is  narrow
(FWHM H$\beta <2000$ km/s) and relatively weak 
(less than a third of the intensity of the [OIII]$\lambda 5007$ line),
and the optical FeII lines are very strong. These 
characteristics locate NLS1s  near  the extreme value of the Boroson and
Green (1992)  Eigenvector 1. 
Soft X-ray  excess and the tendency of the spectrum to flatten at low energy,
have been found in several NLS1s, e.g. 
 PKS 0558-504 (O'Brien et al. 2001), Mrk 766 (Boller et al. 2001), 
PG 1244+026 (Fiore et al. 1998), PG
1404+226, PG 1440+356, PG 1211+143  (George et al. 2000), IRAS 13224-3809
(Vaughan et al. 1999), 1H 0707-495 (Boller et al. 2002, Dewangan et al. 2002),
RX J1702.5+3247 (Gliozzi et  al. 2001) and Ton S180 (Turner et al. 2001a, 
Turner et al. 2002).  
\sax\ and \asca observations of Akn 564 and RE J1034+396 clearly 
show these extreme properties including a soft X-ray hump with a very
flat spectrum ($\nu F_{\nu} = const.$, 
Comastri et al. 2001;  Puchnarewicz et al. 2001; Turner et al.  2001b). Other
unusual X-ray properties are discussed in Brandt et al. (1998), Comastri
et al. (2001), Pounds, Done \& Osborne (1995) and Collinge et al.  (2001). 
Earlier suggestions attributing the soft excess to blends of
emission lines are not supported by recent \asca, \XMM\  and \chandra
observations (e.g. Turner, George \& Netzer 1999; Puchnarewicz et al. 2001,
Turner et al. 2001a, b, Collinge et al. 2001).  

According to some theories, NLS1s contain relatively small (considering their
luminosity)  black holes (BHs) with very high accretion rates (Laor et al.
1994, 1997, Boller et al. 1996) exceeding, in some cases, the critical
Eddington rate (Mineshige et al. 2000, Collin et al. 2002). This hypothesis is
supported by comparing NLS1  spectra with the spectrum of Galactic black hole
candidates (Pounds et al 1995),  and by very few direct measurements of the
BH mass in a handful of NLS1s (e.g. Peterson et al., 2000).
The observed properties of NLS1s, in particular the soft X-ray excess
and the small BH mass, suggest that they may produce much of their emission
due to processes in high  accretion-rate  disks with hot coronae.
It is therefore important to calculate models  of such systems.

\begin{table*}
\footnotesize
\begin{center}
\centerline{ {\bf Table 1.}
Model List of Emergent Spectrum from Accretion Disk with High Accretion Rate}
\vglue-0.3cm
\begin{tabular}{lllccccc}\\ \hline \hline
~~~reference & disk struc.&radiation & Comp.& $z-$structure &GR&corona& note\\
\hline CT95  & global & BB, f-f  &  Yes& No.& No.&No.&shocks(?)\\
SAM96 & global & MBB &No. & No.&No.&No.& \\
WSLZ99& global &f-f, BB&No. & Yes&No.&No.& \\
MKTH00& global &BB &No. &No. & No.&No.& \\ 
WN &self-similar&f-f, BB&Yes&No.&No.&Yes& \\ \hline
\end{tabular}
\vskip 2pt
\parbox{4.8in}
{\small\baselineskip 9pt
\footnotesize
\indent
CT95: Chakrabarti \& Titarchuk (1995); SAM96: Szuszkiewicz, Abramowicz \& Malkan
(1996); WSLZ99: Wang et al. (1999); MKTH00: Mineshige et al. (2000); WN - the
present work\\
 f-f: free-free emission, BB: black body radiation,
MBB: modified black body radiation by
electron scattering, Comp.: Comptonization; 
GR:general relativity.  }
\end{center}
\vglue-0.5cm
\end{table*}
\normalsize

Thin accretion disks are thought to produce the optical-UV continuum
radiation in Broad Line Seyfert 1 galaxies (BLS1s; see 
Shields 1978; Malkan \& Sargent; 1982, Malkan 1983; Laor and Netzer, 1989).
Indeed, thin disk spectra, based on the  standard
disk model (Shakura \& Sunyeav 1973) with Eddington ratio
$L/L_{\rm Edd}<0.3$, give reasonable
fits to the spectrum of many BLS1s (Czerny \& Elvis  1987, Wandel
\& Petrosian 1988, Sun \& Malkan 1989;  Laor \&
Netzer 1989; Laor 1990; Ross, Fabian \& Mineshige 1993, St\"orzer,  Hauschildt
\& Allard 1994; Shimura \& Takahara 1994, D\"orrer et al. 1996;  Sincell \&
Krolik 1997; Hubeny \& Hubeny 1997; Kawaguchi et al. 2001).  However, there
were very few attempts to calculate  the spectrum of high
accretion-rate disks which we call either ``slim disks'' (SDs, after
Abramowicz et al. 1988) or ``extreme slim disks'' (ESDs), depending on their 
accretion rate.  The few models that are available in the
literature are shown in Table 1 where we list their more important ingredients.
Some common properties regarding the  emergent spectrum are the flattening 
at soft X-ray energies, the  high energy cutoff which is almost
independent of the accretion rate, and the fact that the total  luminosity is 
saturated (Wang \& Zhou 1999; Wang  et al. 1999; Fukue 2000; Mineshige et al.
2000). 

Another ingredient in some disk models is a hot corona. If exists, such
a corona must be complex, as is evident from the highly variable X-ray
spectrum. The
 geometry and formation of
the hypothetical  corona are poorly understood  (see a review by Collin 2001).
There are several published models of standard, thin accretion disks with hot
coronae  (see Shapiro et al., 1976; Liang \& Price, 1977, Svensson \& Zdzarski
1994, Janiuk \& Czerny 2000), but, so far,  no theoretical model of
 a high accretion-rate  disk with a  hot corona.

In this paper we study ESDs with hot coronae in order to explain 
the X-ray spectrum of NLS1s.  The paper is arranged as follows:   
\S2 gives the basic equations and explains the method of calculations. 
\S3 discusses the spectral properties of such systems and in \S4 we examine 
the Eddington ratio in several observed systems and apply our model 
to the soft hump observed  in RE J1034+396. Finally, in \S5 we discuss 
implications to future models and observations of NLS1s.

\section{The structure of ESDs with hot coronae}\label{sect:model}
 
\subsection{Extreme slim disks}
The standard accretion disk model (Shakura \& Sunyeav 1973) breaks down
when the accretion rate approaches the Eddington rate. At this limit, the disk
becomes ``slim'' and  radial advection cannot be neglected any more.  A large
range of  accretion rates ($\dotM$) and relative  rates ($\dotm$) of such
systems has been considered (Abramowicz et al.  1988;  Szuszkiewicz et al.
1996, Wang et al. 1999, Mineshige et al. 2000; Watarai \& Mineshige 2001; 
Ohsuga et al. 2002).  Here
\begin{equation} 
\dotm=\frac{\dot{M}}{\dot{M}_{\rm cr}};~~~ 
\dot{M}_{\rm cr}=\frac{64\pi GM_{\rm BH}m_{\rm p}}{\sigma_{\rm T} c}, 
\end{equation} 
$\sigma_{\rm T}$ is the Thomson cross section, 
$m_{\rm p}$ the mass of the proton, $c$ the speed of light, $M_{\rm BH}$ the
BH mass  and we have assumed an accretion conversion efficiency
of $\eta_0=1/16$. ESDs are taken here to be those cases with $2.5\ll
\dot{m}\leq 100$ (see discussions below). In such disks, the emitted photons
are trapped in the gas   due to the  large Thomson  depth which causes 
 the diffusion time to exceed the radial motion time
(Begelman \& Meier 1982; Wang \& Zhou 1999; Spruit 2000; see Ohsuga et al. 2002
for a more sophisticated treatments of photon trapping). In such cases, 
the surface cooling is less efficient with important consequences to the
emergent spectrum. Below  we consider ESDs where advection cooling
dominates over other surface cooling due to such trapping. 

In this work we consider also a hot corona that modifies the emergent
disk spectrum by Compton scattering, and contributes to the high energy
radiation. Previously  published disk-corona models were based on the
following  assumptions: 1) a fraction $f$ of the gravitational energy is
released  in a hot corona and ($1-f$) is deposited in the cold disk (Haardt
\&  Maraschi 1991). 2) the vertical equilibrium  holds in both the
cold disk and the hot corona and is described by the vertically  averaged
equations (Svensson \& Zdzarski 1994, Janiuk \& Czerny 2000). 3) the hot 
corona is cooled by Comptonization of soft photons emitted by the cold disk
including the reflected photons.  With these assumptions, the structure and 
the spectrum of slim disks with hot coronae can be obtained with the 
parameter $f$. There is no way to accurately find the value of $f$ which 
depends on the connection  between the disk and the corona.
We follow a simple prescription and treat $f$ as a free parameter of
the model. The accretion rate in the ESD is thus  $(1-f)\dotM$.
There are  other  constraints on our model, such as the
requirement of an optically thin corona (see below). 

The basic ESD equations are similar to the SD equations for 
a steady disk  in a Newtonian  potential 
of azimuthal symmetry and Keplerian angular velocity   
$\Ome_{\rm k}$.
We take the kinematic coefficient of the
shear viscosity to be $\nu =\alpha a_s \hd$, where $a_s$ is the sound speed,
$\alpha$ the viscosity  parameter, and $\hd$ the half thickness of the  disk. 
The basic equations are those proposed  by 
 Muchotrzeb \& Paczynski (1982), Matsumoto et al. (1984)  and Abramowicz et
al. (1988), except that we neglect the  coefficients $B_i$ (which are of  order
unity)  and the accretion rate is replaced by $(1-f)\dotM$. This gives
\begin{equation}
(1-f)\dm=4\pi R \hd \rhod \vrd,
\end{equation}
\begin{equation}
\frac{\pd}{\rhod}=\hd^2\Ome_{\rm k}^2,
\end{equation}
\begin{equation}
(1-f)\dm (l-l_{\rm in})=4\pi R^2 \hd\alpha \pd,
\end{equation}
\begin{equation}
\frac{1}{\rhod}\frac{d\pd}{dR}-(\Ome^2-\Ome_{\rm k}^2)R+\vrd\frac{d\vrd}{dR}+
\frac{\pd}{\rhod}\frac{d\ln \Ome_{\rm k}}{dR}=0,
\end{equation} 
where $\pd$ is the total pressure, $\rhod$ the 
mass density, $\vrd$ the radial velocity of the flow, $\Omega$ the angular 
velocity, $l$ the specific angular momentum, and $l_{\rm in}$ the eigen value of 
angular momentum at the inner boundary. The last term in equation (5) is 
the correction for the decrease of the radial component of the gravitational 
force away from the equator (Matsumoto et al. 1984). 

While the above equations are suitable to describe the slim disk-corona coupling, 
such calculations are beyond the scope of the present paper. We only consider 
Thomson thin coronae in the limit  $f\ll 1$.  Despite the small $f$, the hot 
corona can be a strong emitter  since $f\dot{M}$ is not necessarily small.

The energy conservation equations should be added to the above equations. 
Advection cooling, which can  dominate over the diffusion cooling, is 
given by 
\begin{equation}
Q_{\rm adv}=-T_{\rm d}\frac{(1-f)\dm}{4\pi R}\frac{dS}{dR}= 
	   \frac{(1-f)\dm}{4\pi R^2}\left(\frac{\pd}{\rhod}\right)\xi,
\end{equation}
(Abramowicz et al. 1988) where $S$ is the entropy, 
\begin{equation}
\xi=(4-3\beta)\gamma_{\rho}-(12-10.5\beta)\gamma_{_T},
\end{equation}
$\gamma_{_{\rho}}=d\ln \rhod/d\ln R$, $\gamma_{_{T}}=d\ln T_{\rm d}/d\ln R$ 
and $\beta$ is the ratio of gas to total pressure.
In this paper we take $\beta=0$ since radiation pressure dominates over gas 
pressure. Thus, $\xi=4\gamma_{\rho}-12\gamma_{_T}$.
For  intermediate values of $f$, the energy equation is
$Q_{\rm adv}+Q_{\rm cool}^{\rm d}=Q_{\rm vis}^{\rm d}$, where 
$Q_{\rm vis}^{\rm d}$  is the energy generation via viscosity in the SD and
$Q_{\rm cool}^{\rm d}$ is the cooling from the surface of the SD.
 For ESDs, 
the condition $f\ll 1$ guarantees that
advection is the  dominant cooling process ($Q_{\rm adv}\gg Q_{\rm cool}^{\rm
d}$) and practically balances 
$Q_{\rm vis}^{\rm d}$, i.e., 
\begin{equation}
Q_{\rm adv}=Q_{\rm vis}^{\rm d}=(1-f)Q_{\rm vis}
\end{equation}
where $Q_{\rm vis}$ is the energy generated by viscosity  per unit area 
\begin{equation}
Q_{\rm vis}=\frac{\dm (l-l_{\rm in})}{4 \pi R}\left(-\frac{d\Ome}{dR}\right)
	   =\frac{n\cj \dm \Ome^2}{4\pi},
\end{equation}
where $n=-d\ln \Ome/d\ln R$ and $\cj=1-l_{\rm in}/l$.
Strictly speaking, the radial motion and energy conservation 
equations can be solved given the appropriate  boundary conditions
(Abramowicz et al. 1988). Denoting $\gamma_p=d\ln \pd/d\ln R$, 
and $\gamma_v=d\ln \vrd/d\ln R$, the  
radial motion equation (5) is reduced to an algebraic equation 
\begin{equation}
\left(\gamma_p-\frac{3}{2}\right)\frac{\pd}{\rhod}-
(\Ome^2-\Ome_k^2)R^2+\gamma_v\left(\vrd \right)^2=0.
\end{equation}
%
Neglecting the boundary condition, i.e., setting
$\cj=1$ (or $l_{\rm in}\approx 0$), and using eqns. 2, 3, 4, 8 and 10, we get, 
after some algebraic manipulations, the following set of
self-similar solutions:
\begin{equation}
\pd=\frac{(1-f)\dm \Ome_k}{4\pi \alpha R},~~
\rhod=\frac{\gamma_0^2}{4\pi \alpha} \frac{(1-f)\dm}{\Ome_k R^3},
\end{equation}
\begin{equation}
\vrd=\alpha \gamma_0^{-1}R\Ome_k,~~
\hd=\gamma_0^{-1}R,
\end{equation}
\begin{equation}
\Ome=\frac{\Ome_k}{\gamma_0},~~
a_s=\frac{R\Ome_k}{\gamma_0},
\end{equation}
where
 $ \gamma_0=(5+\alpha^2/2)^{1/2}$  is a weak function of the
viscosity $\alpha$  which is roughly $\sqrt{5}$ for $\alpha \ll 1$. 
It is interesting to note that $v_r^d, H_d, \Ome$, and $a_s$ are
independent of $f$. When $f$ goes to zero,
the solution reduces to the one given in Wang \& Zhou (1999).

The usual treatment of the inner edge of the accretion disk assumes a certain
last stable orbit, $R_{\rm ms}$, depending on the BH properties  
(e.g. $R_{\rm ms}=3R_s$ for a Schwarzschild
black hole with $R_s=2GM_{\rm BH}/c^2$). This is also taken to be  
 the innermost emitting  radius  $R_{\rm in}$.
Most of the calculations presented here assume a Schwarzschild BH with the
corresponding inner radius. However, Mineshige et al. (2000) and Watarai
\& Mineshige (2001) have shown that a substantial amount of radiation 
can be produced from inside this value of $R_{\rm ms}$ if the accretion rate
is larger than 5$\dotM_{\rm c}$. These authors  take  $R_{\rm in}$
to be  a free parameter of the model. Moreover, it has been  argued 
(Krolik \& Hawley 2002) that 
in real disks, $R_{\rm in}<R_{\rm ms}$, i.e. a substantial fraction of the
observed radiation is emitted inside the last stable orbit. Our calculations
cannot include the Krolik and Hawley effects and we have adopted the
general approach of a free $R_{\rm in}$ as in Mineshige et al. (2000). We
note that   the
smaller   $R_{\rm in}$ is, the better is the self-similar approximation  
since, as explain, for this condition to hold we require  ${\cal J} \simeq 1$.


While surface cooling in ESDs is very small compared with the 
advected energy, the diffuse photons play an
important role  in cooling the corona. We can  calculate  $\Qdcool$
via the  diffusive approximation. The temperature at the central plane of 
an ESD is given by 
$T_d=(3P_d/a)^{1/4}=6.69\times 10^7[(1-f)\dot{m}/m]^{1/4}
\alpha^{-1/4}r^{-5/8}$K, where $a$ is the radiation density constant and
the effective temperature is  
$T_{\rm eff}\approx T_{\rm d}/\tau_{\rm es}^{1/4}$ (Laor \& Netzer 1989). 
For Thomson scattering, $\tau_{\rm es}=\sigma_{\rm T}
\rho_d H_d$, leading to
\begin{equation}
T_{\rm eff}=3.64\times 10^7\gamma_0^{-1/4}m_{\rm BH}^{-1/4}r^{-1/2}{\rm K}, 
\end{equation}
where $m_{\rm BH}=M_{\rm BH}/M_{\odot}$, and $r=R/R_s$. We thus get an
effective temperature distribution of the form
$T_{\rm eff}\propto r^{-1/2}$, as in  Wang \& Zhou 
(1999) and Fukue (2000). This radial dependence 
is caused mostly by photon trapping.
The surface energy flux, $\Qdcool$, is
\begin{equation}
\Qdcool=\sigma T_{\rm eff}^4=1.0\times 10^{26}\gamma_0^{-1}m_{\rm
BH}^{-1}r^{-2}, ~~~{\rm erg~s^{-1}~cm^{-2}}
\end{equation}
We will show later that the black body approximation used here  to calculate
the emergent flux  is justified because of the saturated 
Comptonization in  ESDs.

It is important to note that $T_{\rm eff}$ and $\Qdcool$ are independent 
of the accretion rate and the factor $f$ but  depend on  the 
mass of the black hole and very weakly on the viscosity $\alpha$ through
$\gamma_0$.  Thus we can estimate the  BH mass directly from the
spectrum.

In  ESDs, the time scale of photon diffusion in the $z$-direction is
$\tau_{\rm diff}\sim \hd \kappa_{\rm es}\Sigma/c$ whereas the radial 
transportation time scale of the accreting gas
is $\tau_{\rm R}\sim \pi R^2\Sigma/\dot{M}$ (Mineshige et al. 2000). 
The condition $\tau_{\rm R}=\tau_{\rm diff}$ can be used to obtain the 
dimensionless photon trapping radius, inside which the photons will be trapped
in the gas,  
\begin{equation}
r_{\rm trap}=7.2\times 10^2(1-f)\left(\frac{\dot{m}}{50}\right).
\end{equation}
The total ESD luminosity integrated over  the photon trapping region
is obtained from the integration of equations (15)
\begin{equation}
L_{\rm rad}\approx 8.0\times 10^{37} m_{\rm BH}
            \ln\left[\frac{(1-f)\dot{m}}{50}\right]~~{\rm ergs~s^{-1}}.
\end{equation}
We see that the total radiated luminosity $L_{\rm rad}$ is only weakly
dependent  on the accretion rate $\dot{m}$.  Comparing with the thin disk
critical radius  $r_{\rm h}\approx 14$ (Laor \& Netzer 1989), inside which
half the total luminosity is dissipated, we see that for high accretion rates,
most  of the gravitational energy is released within the trapping radius and
much  of it is advected into the black hole. Because of that, the 
calculations of the  ESD spectrum  neglect the region beyond $r_{\rm trap}$
that contributes  very little to the emergent spectrum.

A necessary condition for our self-similar solution to hold  is 
$Q_{\rm adv}\gg Q_{\rm cool}^{\rm d}$. This is equivalent to $\dot{m}\gg 0.18r$. 
If we demand that advection dominates surface cooling at $r=r_{\rm h}$, namely 
at least half the dissipated energy will be advected, we get $\dot{m}\gg 2.5$ 
which, using eqn. 16 gives $r_{\rm trap}\approx 36$.
As explained, such a high accretion rate leads
to a significant release of gravitational energy inside the last stable orbit. 
Thus, in our model we  allow $R_{\rm in}<3R_{\rm s}$.

\subsection{The hot corona}
We assume that the gas in the optically thin corona rotates with Keplerian 
velocity, and that gas pressure dominates over radiation pressure since
the corona is optically thin.  Advection is mainly due to protons that
are heated close to the virial temperature while the  temperature of the
electrons is much lower due to the inefficient energy exchanges between
electrons and protons and due to Comptonization cooling.  
The assumption that vertical
equilibrium holds means that  the height of the corona ($H_c$) is given by
$H_c=c_s/\Ome_k$, where  $c_s=(P_c/\rho_c)^{1/2}\propto T_p^{1/2}$ is the
sound speed.  The cooling in the corona is determined by Comptonization, and
the  seed  photons are contributed from two sources: 1) direct emission from
the ESD and 2) ESD reprocessing of  photons scattered down by the hot corona. 
Considering  the fraction $\eta$ of the coronal flux directed towards the ESD,
and the  energy averaged albedo $a$, Comptonization cooling can be written as 
\begin{equation}
\Qc=A\left[\Qdcool+\eta(1-a)\Qc\right],
\end{equation}
where $A=\exp(y)-1$ is the Compton amplification factor,  
$y=\tau_{\rm es}\Xi_{\rm e}(1+\Xi_{\rm e})$, and $\Xi_{\rm e}=4kT_e/m_ec^2$. 
Following \.Zycki, Collin-Souffrin \& Czerny (1995), we 
assume
\begin{equation}
\Qc+\Qcadv=\frac{3}{2}\Ome_k\alpha_c P_c H_c,
\end{equation}
where $\alpha_c$ is the viscosity parameter in the corona, which may
be different from the viscosity in the ESD. $\Qcadv$ is the advected energy
due to the protons in the corona,
\begin{equation}
\Qcadv=\frac{f\DM c_s^2}{4\pi R^2}\delta(R),
\end{equation}
where $\delta(R)=d\ln P_c/d\ln R-2.5d\ln T_p/d\ln R$ is a parameter 
controlling the corona advection (Janiuk \& Czerny 2000) that should be 
obtained self-consistently since it is a function of radius (see next 
section). The energy equation in the hot corona can be written as
\begin{equation}
\Qc+\Qcadv=f\Qd.
\end{equation}
%
As explained, in such an optically thin medium, the electron and proton
temperatures can be quite different due to inefficient energy
exchange. Thus, the hot corona is a two-temperature plasma.
The net flux from protons to electrons is given by 
\begin{equation}
\Qc=1.5km_p^{-1}(T_p-T_e)(1+\Xi_{\rm e}^{1/2})\nu_{ei}\rho_cH_c,
\end{equation}
where 
$\nu_{ei}=2.44\times 10^{21}\rho_cT_e^{-1.5}\ln \Lambda$ with
$\ln \Lambda\approx 20$ (Shapiro et al. 1976).
We take $\eta=0.5$ and $a=0.2$ throughout this paper.
The free-free emission is 
$j_{\nu} \approx 6.8\times 10^{-38}n_e^2T_e^{-0.5}$ (we neglect
the Gaunt factor). The ratio of f-f emission to Comptonization 
 obtained from eq.(22) is $5.6\nu_{19}(T_e/T_p)\ll 1$
if $T_e\ll T_p$, where $\nu_{19}=\nu ({\rm Hz})/10^{19}$.
Thus free-free emission can be neglected in the  two-temperature plasma.  
The structure of the hot corona is determined by viscosity heating, and
advection and Comptonization cooling. 

An important  constraint on the value of  $f$ is obtained from  limits on the
Thomson depth of the corona. The number density near the last stable orbit
($R_{\rm ms}$) is  $n_{\rm c}=f\dot{M}/4\pi R_{\rm ms}^2cm_p$, where we
take the radial velocity to be  the speed of light and the height of the 
corona to be $R_{\rm ms}$. The Thomson  depth is thus
$\tau_{\rm c}=n_{\rm c}\sigma_{\rm T}R_{\rm ms}=f\dot{m}/6\eta_0$.
This  means that  $f<6\eta_0\dot{m}^{-1}$ or else the corona is Thomson thick 
(Blandford 1990). This constraint  does not depends on the disk model.
To summarize, the present model is valid  when $\dot{m}\gg 2.5$ and 
$f< 0.0375\left(\dot{m}/10\right)^{-1}$.

\begin{figure*}[t]
\centerline{\includegraphics[angle=-90,width=17cm]{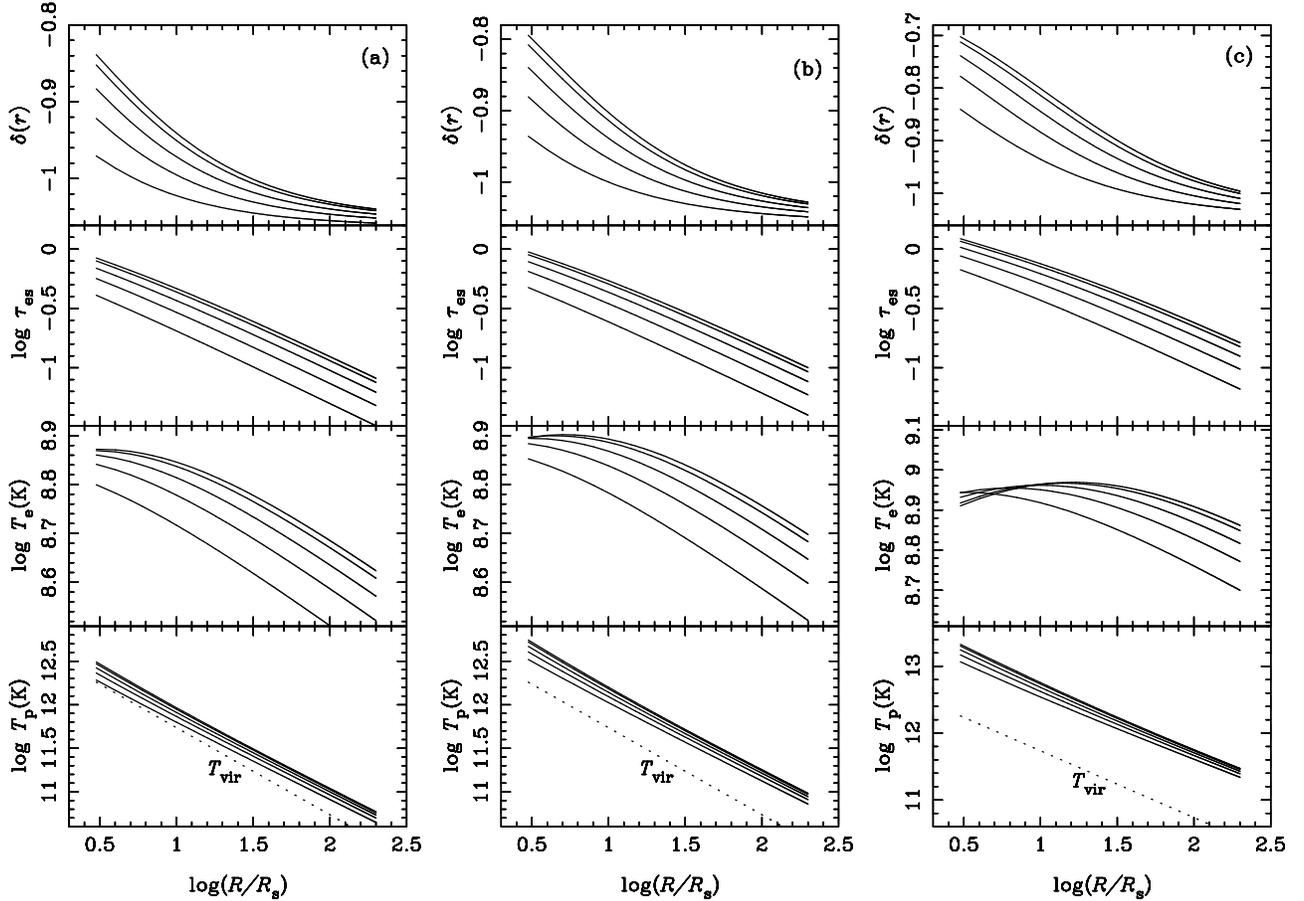}}
\caption{\footnotesize
The structure of hot corona above ESD. 
The accretion rates, from bottom to top, are $\dot{m}=30, 50, 70, 90, 100$
$\dot{M}_{\rm cr}$ 
and $f=10^{-3}$. In Fig-1a $\alpha=\alpha_{\rm c}=0.1$, in Fig 1b
$\alpha_{\rm c}=0.05$ and in Fig 1c $\alpha_{\rm c}=0.01$. 
The advective parameter $\delta (r)$ is obtained from self-consistent
iterations (see text). The hot corona becomes optically thick
when $\dotm>100$ (for $f=10^{-3}$). Under such conditions, emission 
from the disk is  blocked and no hump is observed.
 Using the dimensionless radius $r(=R/R_s)$, the  structure of the corona is
independent of  the mass of the black hole.
\label{fig1}}
\end{figure*}

The structure of ESDs with  hot coronae is controlled by six parameters: 
$\alpha, \dot{m},m_{\rm BH}, \rin$, $\alpha_c$ and $f$. The first four 
describe the ESD and the other two the hot corona. It is important to  
note that the density and the height of the corona depend on the black hole 
mass, but the corona structure, i.e. $T_e$, $T_p$, and $\tau_{\rm es}$, is 
independent of the black hole mass when using the dimensionless radius $r$. 
This implies that the corona 
structure will be similar in stellar and super-massive black hole accreting 
systems. The corona structure is also  only weakly dependent on the accretion 
rate. These may be the reasons why the hard X-ray spectrum of Galactic black 
hole candidates at high state is quite similar to that of narrow line Seyfert 
1 galaxies containing much larger black holes. 

We used an iteration method similar to Chen (1995) and Janiuk \& Czerny 
(2000) to calculate advection in the hot corona. In this procedure,  an 
initial value of advection factor $\delta (R)$  at radius $R$ is assumed. We 
then obtain the structure of the corona at $R-\Delta R/2$ and $R+\Delta R/2$ 
and a new $\delta (R)$ which is  used  for the next iteration until the value 
of $\delta$ converges. In practice, the iteration converges very fast.

Figures 1a-1c show the structure of a hot corona above an ESD for various 
accretion rates $\dotM_{\rm c}=f\dotM$ and viscosity $\alpha$. From 
Fig-1a., we see that the Thomson depth ($\tau_{\rm es}$), electron temperature 
($T_{\rm e}$) and proton temperature ($T_{\rm p}$) increase with $\dot{M}_{\rm
c}$. If we let the factor $f$ increase, the 
corona will gradually become optically thick and blocks the emission 
from the ESD. This gives a constraint on $f$ ($f\ll 1$). 
The temperature of the electrons in the corona is a weak function 
of radius. The optical depth $\tau_{\rm es}$ is a power law
in radius, and is sensitive to the accretion rate $f\dotM$.

We find that the proton's temperature is super-virial,
$T_{\rm vir}=5.45\times 10^{12}r^{-1}~~{\rm K}$.
This temperature is caused by advection, in which
the energy dissipated by viscosity is stored as entropy rather
than being radiated.  Such a process is enhanced 
by the inefficient energy exchange between  electrons and protons.
Advection affects the corona structure, and is found
to be a source of heating rather than cooling. We also see that
$\delta (r)$ tends to a constant at larger radii.

We  tested the effect of changing the viscosity  
on the corona structure (Fig. 1b-1c). We find that 
the scattering depth $\tau_{\rm es}$ increases slightly with decreasing
$\alpha_{\rm c}$ due to the decrease  of radial 
velocity with $\alpha_{\rm c}$, resulting in the increases of
scattering depth. For a hot corona with lower viscosity and higher
accretion rate, we find that the location of 
maximum electron temperature is shifting to very large radii
(e.g.  $\log r\sim 1.5$). These effects have no significant 
 observational consequences.

\section{The spectrum of ESDs with hot coronae}
We have calculated spectra of ESDs with hot coronae over a large range
of BH mass, accretion rate and relative accretion parameter $f$.
Like previous works, we find Comptonization to be very significant in such
systems.  The importance of the process can be estimated by
considering the parameter
\begin{equation}
Y=\Xi_{\rm e}
\left(\frac{\kappa_{\rm es}}{\kappa_{\rm ff}}\right),
\end{equation}
where the Thomson opacity is $\kappa_{\rm es}=0.34$,
 the free-free opacity is 
$\kappa_{\rm ff}=1.5\times 10^{25}\rho T^{-7/2}x^{-3}(1-e^{-x})$,  
and $x=h\nu/kT$. To evaluate $Y$, we use the self-similar solution
(eqs. 11-13). The results are shown in Figure 2 for $\dotm=30$. We find 
$Y\gg 1$ for high energy photons ($x>0.2$) in a large region inside the 
trapping radius, suggesting saturated  Comptonization.  For higher 
accretion rate, $Y$ is even larger, and Comptonization is stronger. Wang 
et al. (1999) and Mineshige et al. (2000) considered the effects of 
advection but neglected Comptonization in their calculations. 
The present paper includes all these processes. Figure 2 shows the
Comptonization parameter $Y$ as a function of the normalized radius.

With high accretion rate, ESDs have unique effective temperature distribution 
of the form $T_{\rm eff}\propto r^{-1/2}$ and relatively high  temperature.  
The emergent spectrum differs significantly from the spectrum of thin 
accretion disks with or without coronae.  We use the method described by 
Czerny \& Elvis (1987) (see also Wandel \& Petrosian 1988) for the spectral 
calculations. The disk structure is self-similar (eqs.11-13) and the surface 
temperature $T_{\rm s}$  is taken to be $T_{\rm eff}$ from equation (14).  The
ESD  spectrum is given by integrating over the photon trapping region, 
\begin{equation}
L_{\rm \nu}=\pi\int_{R_{\rm in}}^{R_{\rm tr}} 
             B_{\nu}(T_{\rm s})f_{\nu} RdR, 
\end{equation}
where the factor $f_{\nu}$ describes the departure 
from black body radiation, including Comptonization,
\begin{equation}
f_{\nu}=f_{\rm ff}(\nu)\left[1-f_{\rm th}(\nu)\right]+C.
\end{equation}
Here the factors $f_{\rm ff}$ and $f_{\rm th}$ are given by 
\begin{equation}
f_{\rm ff}=\frac{2\left[1-e^{-2\tau_{\nu}^*}\right]}{1+
      \left(\kappa_{\rm tot}/\kappa_{\rm abs}\right)^{1/2}};~~
\end{equation}
and
\begin{equation}
f_{\rm th}=\exp \left[-\frac{\ln(kT_s/h\nu)}
  {\tau_{\rm es}^2\ln \left(1+\Xi_{\rm e}+\Xi_{\rm e}^2\right)}
                               \right],
\end{equation}
where $f_{\rm th}$ is the fraction of thermalized photons, 
$\kappa_{\rm tot}=\kappa_{\rm es}+\kappa_{\rm ff}$ is the total  
scattering and free-free opacity, 
$\tau_{\nu}^*=(\kappa_{\rm tot}\kappa_{\rm ff})^{1/2}\Sigma$ 
is the effective optical depth, and the Comptonization factor $C$ is
given in Czerny \& Elvis (1987). 

\begin{figure}
\centerline{\includegraphics[angle=-90,width=8.0cm]{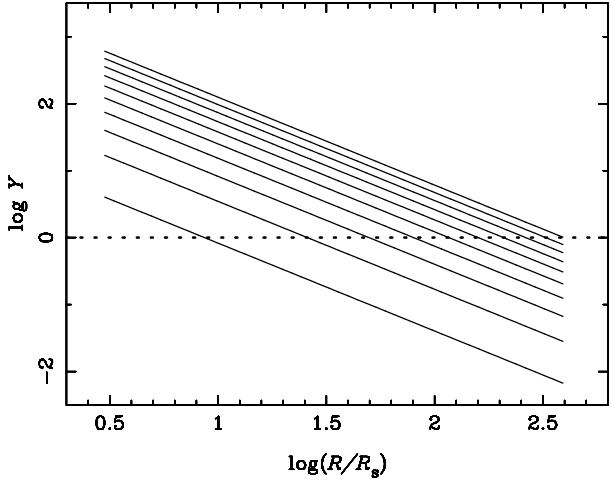}}
\caption{\footnotesize
The Comptonization parameter $Y$ for $\dotm=30.0$ and
$\alpha=0.1$. The solid lines, from bottom to top are:
$x=h\nu/kT$=0.1, 0.2, 0.3, 0.4, 0.5, 0.6, 0.7, 0.8, 0.9, 1.0. 
Comptonization is highly {\it saturated} above the dotted line that indicates
$Y=1.0$.} 
\label{fig2}
\end{figure}

Regarding the hot corona, the emergent spectrum can be 
calculated at each radius from the semi-analytical formula in Hua 
\& Titarchuk (1995). We assume that the photons are Comptonized 
{\it locally}, i.e. we neglect the Comptonization of seed photons 
from other radii. The integration is performed over the trapping radius. 

Figure 3 shows spectra of  ESDs with  hot coronae for different 
accretion rates and black hole mass. The most
prominent feature is a strong soft
X-ray energy band. The luminosity of the hump is insensitive to the accretion 
rate because of photon trapping and  increases only slightly  
as $\sim \ln\dot{m}$ (eq.17).  This results from  
the superposition of {\it saturated} Comptonized spectrum at different
radii, satisfying the condition $Y\gg1$ (Figure 2). For such a large $Y$,
the factor $f_{\rm th}$ tends to $1$ and $C$ tends to a constant (eq. 25).
 With the effective temperature distribution 
$T_{\rm eff}\propto r^{-1/2}$, we get an emergent spectrum
that satisfies the condition $\nu L_{\nu}\propto \nu^{0}$ (eq. 24)
and $L_{\nu}$ is proportional to $m_{\rm BH}$ (see below) . 
This characteristic shape is in rough agreement with the no 
Comptonization case (Wang 1999, Wang et al. 1999, Mineshige et al. 2000),
but the maximum frequency is at higher energy. 

The accretion rate determines the trapping radius which determines also the
shape  of the red side of the soft hump.  Comparing the cases of
$\dot{m}=10$ and  $\dot{m}=40$, we find that the width of the soft hump is
greatly reduced at smaller $\dot{m}$. The behavior can be understood by
inserting eq. (16) into (14) to get the minimum  frequency of the hump
\begin{equation}
\nu_{\rm min}\approx 2.82\times 10^{16}\gamma_0^{-1/4}m_{\rm BH}^{-1/4}
              \left(\frac{\dot{m}}{50}\right)^{-1/2}~~{\rm Hz},
\end{equation}
where we assume $f\approx 0$. The maximum frequency of the hump, which is 
determined by {\it saturated} Comptonization, is independent of $\dot{m}$ and
is given by $\nu_{\rm max} \approx  2.7kT_{\rm eff}(\rin)/h$, namely
\begin{equation}
\nu_{\rm max}
             \approx  1.19\times 10^{18}\gamma_0^{-1/4}m_{\rm BH}^{-1/4}
             \left(\frac{\rin}{3}\right)^{-1/2}~~ {\rm Hz}.
\end{equation} 
The width of the soft hump, from  
$\nu_{\rm min}$ to $\nu_{\rm max}$, depends weakly on the accretion rate.
 This is very different
from  standard thin disk models where the maximum frequency and luminosity
depends on both the BH mass and the accretion rate. The monochromatic
luminosity in the flat part of the hump is not  sensitive to the accretion rate
but depends  linearly on the mass of the  black hole.  This results in  a
good method to determine  a black hole mass for super-Eddington accretion
flows (see below).

The spectrum of the hot corona is determined by two parameters: 
$f\dotm$ and $\alpha_c$. Several examples are shown in Figure 3. 
The general shape is a power-law continuum with a cutoff. The spectrum becomes
harder with increasing $f$. This dependence can be understood with the help of 
Figures 1a-1c that show that the optical depth and temperature of the hot 
corona increase with $f$.  The resulting spectrum can easily be understood by 
the simplest version of Comptonization. The temperature
determines the cutoff energy whereas the photon index $p$ of the spectrum 
is determined by the scattering opacity and the temperature, 
$p=(\gamma+9/4)^{1/2}-1.5$, where 
$\gamma=4\pi^2/3(\tau_{\rm es}+2/3)^2\Xi_{\rm e}$ (Sunyaev \& Titarchuk 
1980). When the accretion rate $\dot{M_{\rm c}}$ (or $f$) increases,
both $\tau_{\rm es}$ and $\Xi_{\rm e}$ increase, resulting in the decrease
of $\gamma$. Thus, the emergent spectrum 
becomes harder as the accretion rate increases.

The overall spectrum shows a "knee", a transiting
from soft to hard X-rays. The flux at the knee frequency
 is contributed from the superposition of the
Wien tail of the soft hump and the Comptonization of the soft hump. The
knee's flux increases with $f$ and with  decreasing viscosity and is
correlated with the hard x-ray spectral index (Fig. 3).  This  is caused by
the fact that the radiation from the  hot corona increases with $f$. 
  Finally, The cutoff  energy of the hot
corona spectrum depends on the maximum temperature of  the electrons. This
cutoff, which is below 1 MeV, may be observed by future  missions such as {\it
INTEGRAL}.

\begin{figure}
\centerline{\includegraphics[angle=-90,width=8.5cm]{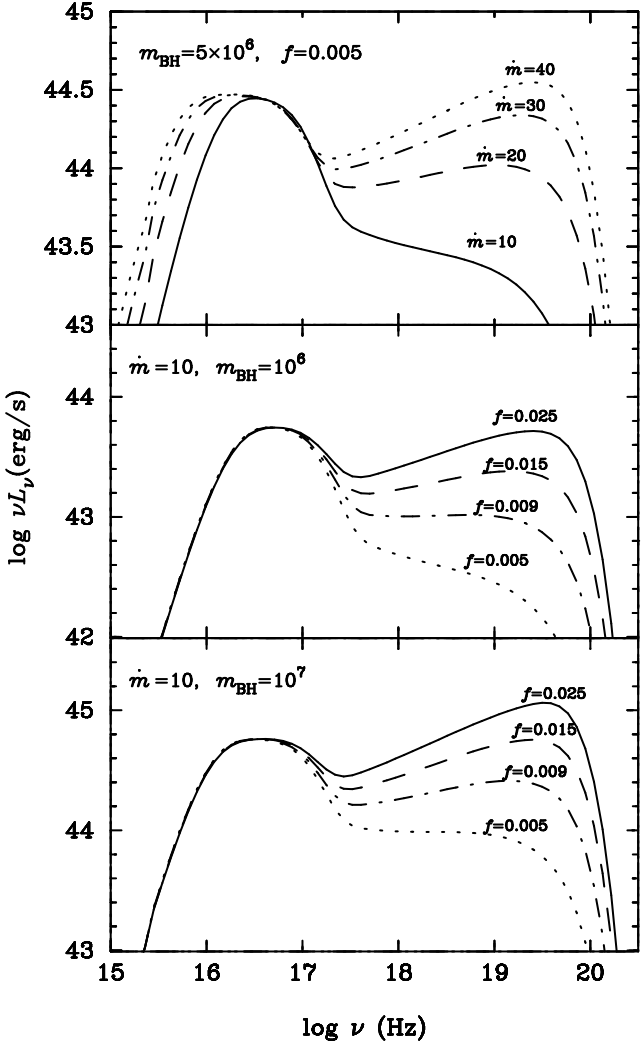}}
\caption{\footnotesize
The emergent spectrum from ESDs with hot coronae for different
accretion rates, black hole masses and $f$. In all cases 
$\alpha=\alpha_{\rm c}=0.1$. The spectrum is composed of two components: 
the ESD contribution at low energy and the hot corona emission at high 
energy. The relative flux of the two depends on the value of $f$ and 
$\dot{m}$ which are marked on the diagram. Note the linear scaling of 
the hump luminosity with $m_{\rm BH}$.} 
\label{fig3}
\end{figure}

The radiation efficiency of an ESD (the fraction
of the total gravitational energy lost to radiation) is low, 
$\eta_{\rm ESD}=L_{\rm rad}/\eta_0\dot{M}c^2\approx 1.6\times
10^{-2}(\dot{m}/50)^{-1}$.  
The  low efficiency implies that most of the photons are
trapped and much of the gravitational energy dissipated by viscosity
is advected into the BH. We call such a flow with
super-Eddington accretion rate ``photon-trapping  accretion flow'' (PTAF).

The low conversion efficiency causes the hump luminosity to be
weakly dependent on the accretion rate. Therefore, even with the uncertainties
on $f$, the  mass of the black hole can be 
estimated from the hump luminosity. Using eq. 14 and integrating
from the inner radius to the trapping radius we get 
\begin{equation}
M_{\rm BH}=2.8\times 10^6\left(\frac{\nu L_{\nu}}{10^{44}{\rm erg/s}}\right)
           M_{\odot},
\end{equation}
where $\nu L_{\nu}$ is the luminosity anywhere in the flat part of the hump,
and we have used the approximation $\int_{x_{\rm in}}^{x_{\rm
trap}}x^3dx/(e^x-1) \approx \int_0^{\infty}x^3dx/(e^x-1)=\pi^4/15$, 
$x_{\rm in}=h\nu/kT(\rin)$, and $x_{\rm trap}=h\nu/kT(r_{\rm trap})$.
This method for determining the BH mass is unique since the  
the exact values of accretion rate and viscosity do not enter the mass
determination. This method does not work  for sub-critical accretion.
 
It is worth pointing out that in the context of standard accretion disks, 
under the assumption that the viscous stress is proportional to radiation 
pressure, the radiation pressure-dominated  region is thermally and viscous
unstable because of the sensitive  temperature dependence on viscosity. 
The optically thick ADAF has no such instability (under the
$\alpha$-description) because the surface
density $\Sigma$ is proportional to $\dotm$ (eqs.11-13). 

\section{Applications to NLS1s}\label{sect:nls1}
\subsection{The Eddington ratio in NLS1s}
There are several claims in the literature regarding the large $L/L_{Edd}$ in
NLS1s (e.g. Pounds, Done \& Osborne 1995; Boller, Brandt \& Fink 1996; 
Laor et al 1997; Collin et al. 2002). However, the available mass estimates 
are remarkably few and, in most cases, highly uncertain. In particular, detailed
reverberation mapping for the purpose of measuring the BH mass, are 
available for only a handful of sources. Several big campaigns, such as the
one described in Collier et al. (2001) and Shemmer et al. (2001), provided
uncertain mass estimates because the observed continuum variability was
very small. Currently there are seven objects with FWHM(H$\beta$)$<1500$
km/sec and measured BH mass. Their
observed properties are listed in Kaspi et al. (2000) and discussed further in
Peterson et al. (2000). According to Peterson et al. (2000), the BH mass in
those sources is considerably smaller than the masses measured in comparable
luminosity BLS1s. This interpretation is still questionable because of the
uncertain BLR size used in those mass estimates.
Collin \& Hur\'e (2001) and Collin et al. (2002) examined  accretion rates
in AGN  using the Kaspi et al. (2000) relationship.
They find that half  the objects in their sample are accreting close to the
Eddington rate or at a super-Eddington rate. The NLS1s in Collin's sample have
the largest Eddington ratio.  

To further test this idea, we have used the compilation of
V\'eron-Cetty et al. (2001) that contains a heterogeneous sample of 54 NLS1s
with FWHM(H$\beta)< 2000$ and $B$-band absolute magnitudes. 
The reverberation relation of Kaspi et al (2000) allows a determination of
$R_{\rm BLR}$ as a function of luminosity over a large range in optical continuum
luminosity. Following Wang \& Lu (2001), we estimated BH masses in this sample 
using the 5100$\AA$ monochromatic luminosity which we obtained by assuming
a ``typical''   optical continuum of the form $L_{\nu}\propto
\nu^{-0.5}$ and  (as in Kaspi et al. 2000) a
bolometric luminosity of  $L_{\rm Bol}\approx 9\lambda L_{\rm
\lambda}(5100\AA)$. Given the measured FWHM, which is taken to be the
one-dimensional gas velocity, we  calculated BH masses assuming virialized
BLR. Combining with the observed flux and redshift, we   estimated the
Eddington ratio for all 54 sources.  The results are  shown in Figure 4. 

Inspection of  Fig. 4 shows that the mean of the sample is 
\begin{equation} 
\left\langle \log \left(\frac{L_{\rm Bol}}{L_{\rm Edd}}\right)\right\rangle
      =0.08\pm 0.31. 
\end{equation}  
i.e,  most of the NLS1 are super-Eddington accretors  (the two extreme cases
are  IRAS 05262+44 and NGC 4051).  Moreover, 
the theoretical results presented earlier,  as well as the results  of
Wang \& Zhou  (1999) and  Ohsuga  et al. (2002),  suggest that in such sources 
the radiated luminosity is significantly lower than the accreted gravitational 
energy  because of  photon trapping.  In particular, in PTAFs,  
$\dot{m}\gg L_{\rm Bol}/L_{\rm Edd}$. It thus seems reasonable to assume that 
most NLS1s are  super-Eddington accretors.

\begin{figure}
\centerline{\includegraphics[angle=-90,width=8.0cm]{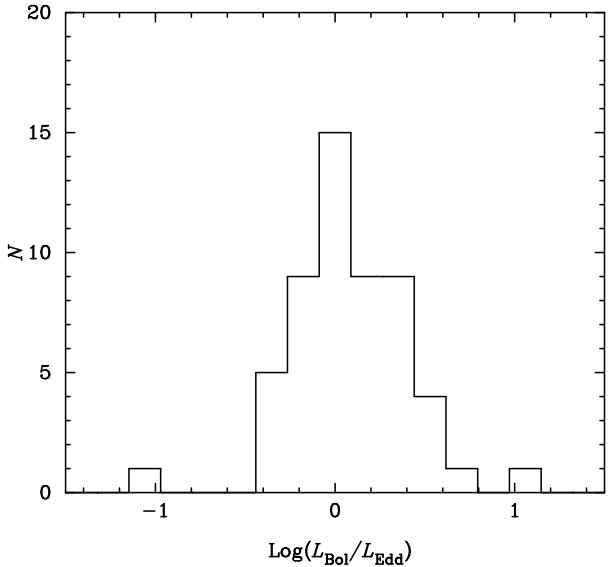}}
\caption{\footnotesize
The Eddington ratio distribution in the  V\'eron-Cetty et al. (2001) sample
calculated as explained in the text.}
\label{fig4}
\end{figure}

As shown in the previous section, super-Eddington accretion can be identified
by the presence of a soft X-ray hump in the spectrum.
The following section addresses the possible observation of such features.

\subsection{Soft X-ray  humps in NLS1s}
Soft X-ray humps in NLS1 spectra have been  discussed in earlier papers 
(Turner, George \& Nandra, 1998; Leighly 1999, Vaughan et al. 1999, 
Comastri et al. 2001; Turner et al. 2001; 
Puchnarewicz et al. 2001;  Collinge et al. 2001; O'Brien et al.
2001; Boller et al. 2001). 
There were several suggestions relating
such features to blends of X-ray emission lines. We consider those unlikely
due to theoretical reasons
(Turner, George  \& Netzer 1999) as well as the lack of direct evidence in
recent high resolution XMM observations of some of the sources. Below we
discuss the spectrum of RE J1034+396 as an extreme case and address 
the possible connection to the theoretical spectra shown in the previous
sections. 

The ultra-soft X-ray radiation of RE J1034+396 ($z=0.003$) 
was first discovered by Puchnarewicz et al. (1995). The spectrum 
shown in Fig. 5 shows a  clear soft X-ray hump extending 
from $\log \nu=16.5$ (the lowest observed frequency) to $17.2$.  
The data were taken from Puchnarewicz et al. (2001). 

The fittings made by Puchnarewicz et al. (2001) to the spectrum of  
RE J1034+396 were based on the standard accretion disk model (Czerny 
\& Elvis 1987). The resulting accretion rate is  0.3--0.7 
$\dotM_{\rm cr}$ and the inclination $\theta_{\rm obs}=60- 75^{\circ}$. 
Such a high accretion  rate is inconsistent with a thin disk model.
For a standard  accretion disk model, the maximum frequency  of 
the spectrum is roughly proportional to $\dotm^{1/4}$
and the luminosity is proportional to $\dotm$. A strong soft X-ray
hump in such disks requires  very large $\dotm$. This would give
very large luminosity which is the reason why Puchnarewicz et al. (2001) had to
assume high inclination (to decreases the total observed  luminosity by 
a factor of $\cos \theta_{\rm obs}$). Even with such extreme  values of
$\dotm$, there were difficulties fitting the observed spectrum.

The ESD model provides a natural way to increase 
$\dotm$ without a proportional increase in the bolometric luminosity.
To illustrate this, we assume a face on system with 
$\alpha=\alpha_{\rm c}=0.1$. The mass of the central black hole can be 
obtained from equation (29).  Fig. 5  shows three such models
fitted to the spectrum of RE J1034+396 representing 
$M_{\rm BH}=2.25\times 10^6M_{\odot}$, 
$\dot{m}=5, 10, 20$ and  $f=1.1\times 10^{-2}, 5.5\times 10^{-3}, 2.75\times
10^{-3}$.  The maximum frequency of the hump is 
determined  by $r_{\rm in}$ which, in this model, is fixed at $r_{\rm in}=3$. 
The minimum frequency  is
determined  by the accretion rate, as seen from the theoretical spectrum in
Fig. 5.  However, in this case the low frequency part of the spectrum 
 is not observed and thus the accretion rate  is
unknown. There are only two parameters, $\bhm$ and $\rin$,  that
determine the total flux  in the hump. The hard X-ray spectrum (at frequencies
larger than $\nu_{\rm max}$) is determined by   $f\dotM$ and
$\alpha_{\rm c}$. Since $\dot{M}$ is unknown,  $f$ is the real free
parameter of the fit.

We find that $\dot{m}=5$ cannot provide a satisfactory fit to the data.
The combination of $\dot{m}=10$ and $f\dot{m}=0.055$ provides a good fit for
the spectrum of and $\dot{m}=20$ is also consistent with the data given the 
lack of observations at low X-ray energies. The large observational 
uncertainties on the  hard X-ray
spectrum translate to large uncertainty on the value of  $f$.
The best fit is  $\bhm=2.25\times 10^6\sunm$,
$f\dotM=0.055\dotM_{\rm cr}$,  $\dotM=10\dotM_{\rm cr}$ and  $\rin=3.0$. 
We have also computed a model with $\rin=2.0$ and found very little change
thus the dependence on $\rin$ over the fitted energy range is very weak.

\begin{figure}
\centerline{\includegraphics[angle=-90,width=8.5cm]{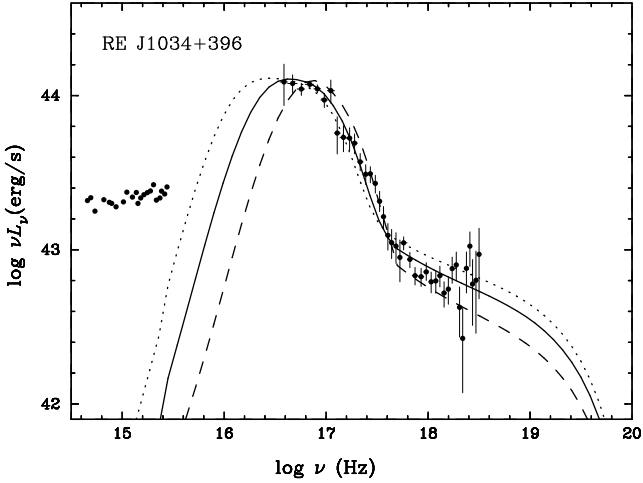}}
\caption{\footnotesize
The soft hump and hard X-ray spectrum of RE J1034+396.
The data are taken from Puchnarewicz et al. (2001).
The dashed, solid and dotted lines represent ESD models with  $\dot{m}=5, 10$,
and 20 respectively, and $f\dot{m}=0.03$. In all cases  
$r_{\rm in}=3$ and $M_{\rm BH}=2.25\times 10^6M_{\odot}$. 
The fits include Comptonization in both the ESD and the hot corona. }
\label{fig5}
\end{figure}

 The BH mass
obtained from the model can be compared with the mass obtained  by using the
Kaspi et al. (2000) relationship for $R_{\rm BLR}$ vs $L_{5100}$, and  the
observed line widths. Using the data in  Puchnarewicz et al. (1995) with 
FWHM(H$\beta$)=1500 Km/s, we get $M=4.1\times 10^6\sunm$, in good agreement
with the spectra-fit results.

The present model allows two different  kinds of X-ray variations. The hard
X-ray spectrum  is the Comptonized component of the soft hump flux. 
This means that intrinsic variations in the hump luminosity must result in
correlated hard X-ray variations but not necessarily in 
optical-UV variations.  On the other hand, the hard X-ray luminosity is very
sensitive to the value of 
 $f$. Time dependent variations  in  $f$ will lead to variations in the
hard X-ray flux yet  the soft hump canremain
constant because of photon
trapping effects. This can be tested observationally. 

\section{Discussion and Conclusions}
In this paper we explored the structure and the  spectrum of
extreme slim disks with hot coronae. We focused on the question of whether
such systems  can fit, simultaneously,  the soft X-ray hump and the hard
X-ray spectrum of NLS1s.  We presented  a self-similar solution 
of ESDs with hot coronae and showed that their spectrum contains
 low energy humps produced by the  
ESD and a higher energy power-law continuum with a cutoff produced by  the hot
corona. We find that the structure of the  hot corona, which we assumed to be
Thomson-thin,  is independent of  the mass of the black hole and is fully
specified by the parameters $f\dot{m}$ and   $\alpha_c$. 
The ESD structure and spectrum (the ``hump'') is  insensitive to the
accretion rate, but  sensitive to the mass of the black hole. This can explain
 the  similarity of the hard X-ray spectrum of Galactic black hole candidates 
and NLS1s despite of the very large difference in BH mass.
Our present models apply to  $\dot{m}\gg 2.5$ and
$f<0.0375\left(\dot{m}/10\right)^{-1}$. The calculations do not include GR
effect and apply  only to non-rotating BHs.

 Slim disk fluctuations have been suggested by
Mineshige et al. (2000) as the origin of the soft X-ray variations in such
systems. The reason is the unusual energy balace 
\begin{equation}
 u_{\rm mag}\sim u_{\rm
grav}>u_{\rm rad}
\end{equation}
 where $u$ denotes the energy density.
The trapped
photon accretion flows discussed in this work are likely to show 
 two different types of variations: 
1) simultaneous variations
of the soft hump and the hard X-ray without any significant time lag; 2) small
variability of the soft hump and  larger variation in the hard X-ray spectrum.
The first type is related to the  so called ``photon bubble instability''
(Gammie 1999) that can  be very important in SDs since the energy densities of
the trapped photons and the magnetic field are larger than those in standard
accretion disks. In this case, the hard X-rays will follow closely the soft
hump variations.
 Using the simplest version of Comptonization (Sunyaev \&
Titarchuk 1980) we find that the ratio 
of the hard X-ray flux to
the soft hump flux is roughly 
$p(p+3)(2p+3)^{-1}\left(\nu_{\rm HX}/\nu_{\rm
hump}\right)^{-p}$ 
where $\nu_{\rm HX}$ and $\nu_{\rm hump}$ are representative frequencies
of the hard and the soft (hump) X-ray flux. This ratio is  sensitive to the
hard X-ray spectrap index  $p$.  This type of variations agrees  with at
least some observations (e.g. Leighly 1999, Dewangan et al. 2002).
Regarding the second type,
here  the fluctuations are  due to changes in the accretion rate.
Such fluctuations cause large hard X-ray variations since that radiation
is sensitive to the accretion. On the other hand,  the soft
hump flux remains almost constant  due to photon trapping.  This type of variations 
has been reported in  the {\it Chandra}
observations of 1H 0707-495 (Leighly et al.  2002). It will be interesting to
apply a more detailed combined spectral and stability analysis of this type to
the observed X-ray spectrum of NLS1s.

The calculations presented here include the new ingredients of a hot corona,
 yet they are still oversimplified in various  ways.
First, we assumed a continuous  corona  which is
characterized by a single parameter $f$. However, patchy corona may be more reaslistic
for  SDs. Second, the hot corona gas is assumed to rotate with
Keplerian velocity. This leads to an underestimate of the advection in
the hot corona since the hot gas may be sub-Keplerian. This uncertainty can be
aborbed into the factor $f$. A possible improvement is to use the
optically thin advection-dominated self-similar solution (Narayan \& Yi 1994)
as an approximation of the hot corona. We also note that for the case of standard 
accretion disk, the factor $f$ approaches unity (Svensson \& Zdzarski 1994).
However, in that case  feedback to the cold disk can be neglectable since
$f\ll 1$ in the super-Eddington case. Third, we do not take into account 
the irradiation of the disk due to the hot corona.  Future work
will have to treat the more general case of  irradiation and try to solve,
self-consistently, for the disk-corona interface e.g. via
ionization instability (\.Zycki, Collin-Souffrin \& Czerny 2000).

Another limitation of the model is the neglect of GR effects. Thus, the
treatment of the physical processes near the disk inner boundary, and the
calculation of the emitted spectrum from this region, are not very accurate.
As explained, global energy considerations show that the models
presented here do not deviate much from the more realistic cases since a large
fraction of the radiation is emitted way outside of  $R_{\rm in}=3$, our
standard inner radius. In general we expect models with full GR treatment and
similar $\dot{m}$ to show a very similar hump and to require a somewhat
smaller $f$ for the same hard X-ray flux. We will test these qualitative
estimates in a future paper. Finally, we  neglected any spectral dependence on
the inclination angle of the disk. This will be included in  future models.

Our new model enables a comparison with several already observed NLS1s.
Study
of the  V\'eron-Cetty et al. (2001) sample showed that the mean Eddington
ratio in this sample is larger than unity which suggests super-Eddington
accretion rates in many NLS1. As an application of the model, we fitted the
spectrum of RE J1034+396 and obtained a good agreement, within the
observational uncertainties, over the entire observed X-ray band.  We also
find that the mass of the central black  hole obtained from the model is in
good  agreement with the mass obtained by using the reverberation mapping
method.

Regarding new observations to test the model, perhaps the most important
are {\it Chandra} and {\it XMM} spectra that cover the lowest energy
range. This will enable a careful study of the ``red'' side of the ESD
hump in several sources and a more direct determination of the accretion
rate.
A careful study of the ``knee'' spectral region is another observational
challenge with implications to the type of black hole (via the value
of $R_{\rm in}$) and the energy production in the corona (via the value of
$f$).

\begin{acknowledgements}
We are grateful to the referee for the useful  comments and suggestions that
helped improving this paper. We also 
  thanks  A. Laor for his careful reading
of an early version of the manuscript and for providing useful comments. We
 acknowledge  useful discussions with  O. Shemmer, D. Chelouche and S.
Mineshige. This
research was supported by the Israel Science Foundation (grant no. 545/00)
and by the Special Funds for Major State Basic Research Projects and
NSFC.  JMW is supported
by Hundred Talents Program of Chinese Academy of Sciences. 
\end{acknowledgements}

\end{document}